\documentclass[%
 reprint,
 amsmath,amssymb,amsrm
 aps,
]{revtex4-2}

\usepackage{graphicx}
\usepackage{dcolumn}
\usepackage{bm}
\usepackage{amsmath}
\usepackage{braket}
\usepackage{algorithm}
\usepackage{algorithmic}
\usepackage{color}

\usepackage[justification=RaggedRight]{caption}
\usepackage[subrefformat=parens]{subcaption}
\newcommand{\tr}{{\rm tr}}

\begin{document}


\title{Quantum statistical effect induced through conditioned post-processing procedures with unitary $t$-designs}

\author{Hideaki Hakoshima}
\email{hakoshima.hideaki.qiqb@osaka-u.ac.jp}
\author{Tsubasa Ichikawa}%
 \email{ichikawa.tsubasa.qiqb@osaka-u.ac.jp}
\affiliation{%
Center for Quantum Information and Quantum Biology, Osaka University, Osaka 560-0043, Japan
}%


\date{\today}

\begin{abstract}
We propose a few-body quantum phenomenon, which manifests itself through stochastic state preparations and measurements followed by a conditioned post-processing procedure.
We show two experimental protocols to implement these phenomena with existing quantum computers, and examine their feasibility by using simulations.
Our simulation results suggest that the experimental demonstration is feasible if we repeat the state preparations and measurements about thirty thousand times to three-qubit systems.
\end{abstract}

\maketitle


\section{Introduction}
\label{sec:level1}
The coin toss game is a typical problem in probability theory.
Consider the following question:
If we observe $k$ heads in the $n$ tosses, what is the probability that we will observe the head at the $(n+1)$-th time coin toss?

If we use the empirical probability distribution (the relative frequency of the events occurring), then the answer is $k/n$.
This is a reasonable answer when $n$ and $k$ are large enough, but we come up against a difficulty when $n$ is small. 
For example, the relative frequency of finding the head up is either unity or zero for any coin when $n=1$, implying that we are forced to accept a counter-intuitive conclusion: any coin is completely biased.
This problem inherent in the case of the small number of observations has been known as the zero-frequency problem in the machine learning community \cite{Witten1991}.

For a resolution of the zero-frequency problem, we may introducing the pseudo counts \cite{Witten1991}; we define the conditional probability $p_{n,k}^{\rm cl}$ of finding the head given the $k$ heads in the $n$ tosses by
\begin{equation}
    p_{n,k}^{\rm cl}=\frac{k+1}{n+2},
    \label{pcl}
\end{equation}
which is called the Laplace law of succession (LLS) \cite{Laplace02, Gillies00, deFinetti37, deFinetti92,jaynes03}, or Laplace smoothing \cite{manning08retrieval}.
Intuitively, this expression operationally means that we toss the coin twice prior to the game, and ensure that we actually observe both the head and tail.
We should evaluate the probability by taking these tosses into account.
The pseudo count is the two tosses performed before the game.
Note that $p_{n,k}^{\rm cl}$ approaches the relative frequency in large $n$ and $k$.

The Laplace law of succession was derived on the ground of the Bayesian considerations.
We are supposed to be ignorant of how much the coin is biased, and thereby forced to introduce the uniform distribution of the bias.
We can then obtain LLS by averaging the conditional probability under a given unknown fixed bias over the uniform distribution.
Note that we may perform the integration with a non-uniform prior if we have reasonable evidence \cite{Laplace02,jaynes03}.
In such case, the pseudo counts in general depend on what prior distribution we use. 
For example, we can obtain $\frac{k+\lambda^2/2}{n+\lambda^2}$ if we take into account information on the standard deviation, where $\lambda$ is a free parameter to express \lq\lq our readiness to the gamble on the typicalness of our realized experience\rq\rq \cite{10.2307/2276774}.
Then the pseudo count $\lambda^2$ is a formal correction for the relative frequency, and we are not necessarily able to find the clear operational interpretation thereof. 

Recently, LLS is generalized to quantum mechanics (QM) \cite{Ichikawa2021} as a consequence of the Bayesian derivation of quantum (conditional) probability \cite{Caves02, Fuchs13, Ichikawa2021}.
Hereafter we call it a quantum Laplace law of succession (QLLS).
In this generalization, our ignorance is formulated by an appropriate distribution of quantum states. 
Since the state space (Hilbert space for the pure states) has degrees of freedom associated with the unitary transformations, $p_{n,k}^{\rm cl}$ acquires a non-trivial overall factor, which approaches unity in the large $n$ and $k$.
Thus, similarly to $p_{n,k}^{\rm cl}$, QLLS also approaches the relative frequency, but its overall factor manifests itself at a small number of the observations.

So far, we have followed the Bayesian viewpoint and regarded the uniform distribution introduced for the derivation as our prior information on how much the coins are biased.
In parallel, we can derive LLS by another approach: 
we draw a number $w\in[0,1]$ from an random number generator, 
whose outcome $w$ is the bias in the coin.
In this setup, we observe LLS by repeating these drawings and trials many times and evaluating the conditional probability that we observe the head of the $(n+1)$-th coin, given the $k$ heads of the other $n$ coins.

As to be shown in the following section, this setup is readily extended to QM, and we can derive QLLS.
In this non-Bayesian setup, QLLS is a statistical formula of quantum conditional probability, and, in comparison with the (classical) LLS, its non-trivial overall factor can be regarded as a characteristic feature of quantum few-body systems; one may expect its experimental demonstration.

For the experimental demonstration, it is of relevance that how to implement the distribution of the quantum states.
Since the uniform distribution of the quantum states should have non-vanishing support for almost all of the quantum states, the sampling from the distribution requires large numbers of trials.
We need to replace the distribution with a more tractable one, with QLLS being virtually left unchanged.

In this paper, we propose experimental protocols for QLLS by overcoming the difficulty mentioned above.
The point is the use of the unitary $t$-design~\cite{PhysRevA.90.030303,scott2008optimizing,PhysRevA.80.012304,roberts2017chaos,PRXQuantum.2.030339}, which is a set of {\it finite} numbers of unitary operators and emulates the behavior of the Haar random unitary distribution up to the $t$-th moment of a given observable.
Therefore, the sampling from the $t$-design requires finite numbers of trials and the use of it does not affect QLLS, as to be shown in our simulation results.
We shall also show that the outcomes of our experimental protocols converges to the theoretical predictions in accordance with the central limit theorem.

The rest of the paper is organized as follows:
In Sec.~\ref{th}, we give a brief review of the derivation of QLLS.
In Sec.~\ref{imp}, we propose two experimental protocols of QLLS by making use of the unitary $t$-design.
Through numerical simulations, we demonstrate that these protocols work well in Sec.~\ref{sim}.
Section \ref{conc} is devoted to our conclusion.


\section{Theory}
\label{th}
\subsection{Setups}
Let us formulate the coin-toss problem posed in Introduction by using QM. 
%
Suppose that our coins are $(n+1)$ two-level systems (qubits), all of which are mutually distinguishable and prepared to a density matrix $\rho$, which acts on the two-dimensional Hilbert space $\mathbb{C}^2$ and corresponds with the bias of the classical coin. We hereafter denote the Hilbert space of the $i$-th qubit by ${\cal H}_i=\mathbb{C}^2$ for $i=1,2,\dots,n+1$. Given a probability measure $\mu(\rho)$ of the density matrices, we define the density matrix $\rho_{n+1}$ of the $(n+1)$-qubit system as
\begin{equation}
    \rho_{n+1}=\int d\mu(\rho)\rho^{\otimes(n+1)}.
    \label{def_rho}
\end{equation}

The state $\rho_{n+1}$ is a fully separable mixed state for any probability measure $d\mu(\rho)$.
In contrast, as shown in Appendix \ref{discord_cal}, $\rho_2$ with the flat measure to be defined later exhibits {\it non-zero} (global) quantum discord \cite{PhysRevA.84.042109, xu2013analytical,Bera_2018}, which is a yardstick of a non-classical correlation in terms of the difference between the quantum relative entropy for the total system and those of the subsystems, both of which are associated with local projection measurements.
This suggests that 
$\rho_{n+1}$
may also have non-zero quantum discord in general and we could expect non-classical behavior in the quantum coin-toss problem.

The coin toss with respect to a qubit is performed by the projection measurement along the computational basis $\{|0\rangle, |1\rangle\}$, where
$
|0\rangle=(\begin{smallmatrix}
1\\
0
\end{smallmatrix})
$
and
$
|1\rangle=(\begin{smallmatrix}
0\\
1
\end{smallmatrix})
$ correspond with the head and tail of the coin, respectively.
Accordingly, we introduce a projector $P=|0\rangle\langle0|$ and its orthogonal complement $\lnot P=\openone-P=|1\rangle\langle1|$, where $\openone$ is the identity operator acting on $\mathbb{C}^2$, and write the probability of finding the head by
\begin{equation}
    p=\tr(\rho P).
\end{equation}
Note that $\rho$ is a single-qubit density matrix, different from $\rho_{n+1}$ in Eq.~(\ref{def_rho}).

We further introduce the projector describing the measurement outcomes of the $k$ heads in a toss of $n$ coins as
\begin{equation}
    \Pi_{n,k}=P_{n,k}+({\rm permutations}),
    \label{perm}
\end{equation}
where
\begin{equation}
    P_{n,k}=P^{\otimes k}\otimes(\lnot P)^{\otimes (n-k)}
\end{equation}
and the sum in the RHS of Eq.~(\ref{perm}) is taken over all the $n$-partite tensor product projectors obtained by the permutation of $P$ and $\lnot P$ in $P_{n,k}$. 
Experimentally, $\Pi_{n,k}$ corresponds to the measurements of $\{P, \lnot P\}^{\otimes n}$ for the $n$ qubits, followed by the aggregation of all the outcomes having the $k$ heads.
Given $\Pi_{n,k}$, using the formula of the quantum conditional probability
\begin{equation}
    p(P|Q)=\frac{\tr(Q\rho QP)}{\tr(\rho Q)}
\end{equation}
for the projectors $P,Q$ and the density operator $\rho$, we write the conditional probability as
\begin{equation}
    p_{n,k}
    =p(P\otimes\openone^{\otimes n}|\openone\otimes \Pi_{n,k})
    =\frac{\tr(\rho_{n+1}P_{n+1,k+1})}{\tr(\rho_nP_{n,k})},
    \label{q_cond}
\end{equation}
for the measurements on $\rho_{n+1}$.
The density matrix $\rho_n$ appears in the denominator of the RHS of Eq.~(\ref{q_cond}), since $\tr(\rho\otimes (\rho^{\otimes n}\Pi_{n,k}))=\tr(\rho^{\otimes n}\Pi_{n,k}))$) and the exchange symmetry in $\rho^{\otimes n}$ allows us to replace $\Pi_{n,k}$ with $P_{n,k}$ up to the overall combinatorial factor which cancels out with that coming from the numerator.
See \cite{Ichikawa2021} for the details of the derivation.

\subsection{Quantum Laplace law of succession}
\label{QLLS}
To derive QLLS, let us first perform the spectral decomposition 
\begin{equation}
    \rho=U\Lambda U^\dagger,
    \quad
    \Lambda=\lambda|\psi\rangle\langle\psi|+(1-\lambda)|\psi^\bot\rangle\langle\psi^\bot|
    \label{dec}
\end{equation}
with use of $0\le\lambda\le1$, 
an orthonormal basis $\{\ket{\psi}, \ket{\psi^\bot}\}$
and $U\in SU(2)$, and introduce the product measure
\begin{equation}
    d\mu(\rho)=df(\lambda) d\nu(U),
    \label{product}
\end{equation}
which is the product of the measure $df(\lambda)$ of the eigenvalue $\lambda$ and the measure $d\nu(U)$ of the $SU(2)$ group elements $U$.

Let us mention an interesting identity by restricting ourselves to consider the measure $df(\lambda)$ satisfying the following symmetry
\begin{equation}
    f(\lambda)=f(1-\lambda).
    \label{f_sym}
\end{equation}
This symmetry leads to a duality relation
\begin{equation}
    p_{n,k}+p_{n,n-k}=1,
    \label{reciprocal}
\end{equation}
which implies
\begin{equation}
    p_{2n,n}=\frac{1}{2}.
    \label{half}
\end{equation}
For the proof of Eq.~(\ref{reciprocal}), see Appendix \ref{proof_reciprocal}.
We shall see that these relations approximately hold even in the numerical examples, where statistical errors may affect them.

\begin{table}[b]
    \centering
    \caption{Conditional probabilities for the (quantum) Laplace law of succession. $p_{n,k}\neq p_{n,k}^{\rm cl}$ implies the non-classical feature of $p_{n,k}$. The duality relations (\ref{reciprocal}) and (\ref{half}) due to the symmetry of the measure are ensured.}
    \label{values}
\begin{tabular}{c|ccc}
\hline
    & $p_{n,k}^{\rm cl}$  & \multicolumn{2}{c}{$p_{n,k}$} \\
    measure   & uniform & flat & Bures \\
    \hline
    \hline\\
    $(n,k)=(2,0)$& 0.25 & 0.40 & 0.30\\
    $(n,k)=(2,1)$& 0.50 & 0.50 & 0.50\\
    $(n,k)=(2,2)$& 0.75 & 0.60 & 0.70\\
    &&&\\
    \hline\\
    $(n,k)=(4,0)$& 0.17 & 0.32 & 0.21\\
    $(n,k)=(4,1)$& 0.33 & 0.43 & 0.36 \\
    $(n,k)=(4,2)$& 0.50 & 0.50 & 0.50 \\
    $(n,k)=(4,3)$& 0.67 & 0.57 & 0.64\\
    $(n,k)=(4,4)$& 0.83 & 0.68 & 0.79\\
    &&&\\
\hline
\end{tabular}
\end{table}

Let us focus on more symmetric case such that
\begin{equation}
    d\nu(U)=dU,
\end{equation}
where $dU$ is the Haar measure.
In this case, from the symmetry, we may hereafter work with 
\begin{equation}
    \ket{\psi}=\ket{0},
    \qquad
    \ket{\psi^\bot}=\ket{1}
    \label{redef}
\end{equation}
in the calculation of the density matrix (\ref{def_rho}).
The conditional probability $p_{n,k}$ takes the form of
\begin{equation}
    p_{n,k}=\frac{k+1}{n+2}\frac{I_{n+1,k+1}}{I_{n,k}},
    \label{qLL}
\end{equation}
for the flat measure
\begin{equation}
    df(\lambda)=d\lambda
    \label{flat}
\end{equation}
with
\begin{equation}
    I_{n,k}=\sum_{j=0}^k\sum_{l=0}^{n-k}\binom{j+l}{j}\binom{n-j-l}{k-j}B(n-r+1, r+1),
\end{equation}
and the Bures measure
\begin{equation}
    df(\lambda)=\frac{2}{\pi}\frac{(2\lambda-1)^2}{\sqrt{\lambda(1-\lambda)}}d\lambda,
    \label{Bures}
\end{equation}
with
\begin{eqnarray}
    I_{n,k}&=&\frac{2}{\pi}\sum_{j=0}^k\sum_{l=0}^{n-k}\binom{j+l}{j}\binom{n-j-l}{k-j}\nonumber\\
&&\times\frac{(n-2r)^2+n+1}{(r+\frac{1}{2})(n-r+\frac{1}{2})}B(n-r+\frac{3}{2}, r+\frac{3}{2}),
\end{eqnarray}
where $B(n,k)$ is the Beta function and we have introduced
\begin{equation}
r=k-j+l.    
\end{equation}
The explicit forms of $I_{n,k}$ can be derived by the direct calculation:
in the case of qubits, we may employ the Euler angle representation for $SU(2)$ elements. Using this, we expand the product of the traces coming from the tensor product in Eq.~(\ref{def_rho}) into the power series, and perform the integral.
See \cite{Ichikawa2021} for details of the derivation.

%

It has been shown for both cases that $I_{n+1,k+1}/I_{n,k}$ approaches 1 in the large $n$ limit, which ensures $p_{n,k}$ approaches the classical counterpart $p_{n,k}^{\rm cl}$.
On the other hand, unlike the classical case, $I_{n+1,k+1}/I_{n,k}\neq1$ when the qubit number $n$ is small such as typically ${\cal O}(1)$.
This difference is due to the fact that the flat measure (\ref{flat}) and Bures measure (\ref{Bures}) are the products of the classical measure and the Haar measure; we may see the behavior of $p_{n,k}$ for small $n$ as a quantum mechanical phenomenon coming from the degrees of freedom pertinent to the unitary transformations.

Moreover, as shown in Appendix \ref{derivation_cl}, QLLS can be reduced to LLS when we choose
\begin{equation}
    df(\lambda)=d\lambda,
    \qquad
    d\nu(U)=\delta(U-\openone)dU,
    \label{classical}
\end{equation}
with the choice (\ref{redef})
, where $\delta(x)$ is the Dirac delta function.
This clearly supports the aforementioned scenario that the integral over the unitary transformations contributes to the non-trivial overall factor in QLLS.

All the measures considered so far satisfy the symmetry (\ref{f_sym}), and we may expect the relations (\ref{reciprocal}) and (\ref{half}).
Table \ref{values} shows the actual values of the conditional probabilities $p_{n,k}^{\rm cl}$ and $p_{n,k}$ for $n=2,4$, where the announced relations clearly appear.
Moreover, we immediately observe $p_{n,k}\neq p_{n,k}^{\rm cl}$ except the case $n=2k$, implying the contribution from the overall factor $I_{n+1,k+1}/I_{n,k}$.

\section{Implementations}
\label{imp}
Although QLLS is a natural extension of its classical counterpart, we confront a difficulty if we wish to perform its experimental verification.
This is owing to the fact that the Haar measure $dU$ and $df(\lambda)$ have non-vanishing supports almost everywhere of their domains in their respective parameter spaces.
Since the parameters are {\it continuous}, the implementation of QLLS requires large numbers of sampling if we wish to perform (unbiased) sampling over a sufficiently large subset in the parameter space. 

In this section, we first propose how to reduce the number of the sampling.
We will use the unitary $t$-design~\cite{PhysRevA.80.012304,scott2008optimizing,PhysRevA.80.012304,roberts2017chaos,PRXQuantum.2.030339} and discretize the integral over $\lambda$.
We will next show the experimental procedure and post-processing to evaluate QLLS.

\subsection{Reduction of the unitary degree of freedom by using unitary $t$-design}

The unitary $t$-design ${\cal C}_t$ is a set of $SU(2)$ unitary operators, which reproduce the $t$-th moment of the observable $O$ over the Haar measure:
\begin{equation}
    \int dU U^{\otimes t}O(U^\dagger)^{\otimes t}
    =\frac{1}{|{\cal C}_t|}\sum_{U\in{\cal C}_t} U^{\otimes t}O(U^\dagger)^{\otimes t}.
\end{equation}
Then, by using the spectral decomposition (\ref{dec}), the state $\rho_{n}$ for two measures is rewritten as
\begin{equation}
    \rho_{n}=\frac{1}{|{\cal C}_{n}|}\int_0^1df(\lambda)\sum_{U\in{\cal C}_{n}} (U\Lambda U^\dagger)^{\otimes n},
    \label{rho_cn}
\end{equation}
where $|{\cal C}_{n}|$ is the cardinality of the unitary $n$-design ${\cal C}_n$.
Since the cardinality of the $n$-design is finite, it suffices to sample $U$ from ${\cal C}_n$ finite times in order to create $\rho_n$.
Moreover,
%
by substituting Eq.~(\ref{rho_cn}) into RHS of Eq.~(\ref{q_cond}), we find the closed expression of $p_{n,k}$ in terms of $t$-design as
\begin{equation}
    p_{n,k}=\frac{|{\cal P}_n|\int_0^1df(\lambda)\sum_{\Pi\in{\cal P}_{n+1}}\tr(\Lambda\Pi)^{k+1}\tr(\Lambda(\lnot \Pi))^{(n-k)}}{|{\cal P}_{n+1}|\int_0^1df(\lambda)\sum_{\Pi\in{\cal P}_n}\tr(\Lambda\Pi)^k\tr(\Lambda(\lnot \Pi))^{(n-k)}},
    \label{p_t_design}
\end{equation}
where
\begin{equation}
    {\cal P}_n=\{\Pi=UPU^\dagger\,|\, U\in{\cal C}_n \}
\end{equation}
is the set of the projectors associated with the unitary $n$-design.

Two remarks are in order:
First, Eq.~(\ref{p_t_design}) shows that $p_{n,k}$ can be emulated if we are given $f(\lambda)$ and $\tr(\Lambda\Pi)$ for all $\Pi\in {\cal P}_n$.
This implies that we need only sampling of the expectation values followed by an appropriate post-processing, in order to evaluate $p_{n,k}$.
Second, many other unitary groups satisfying unitary $t$-design are known in the previous studies.
Particularly in a single-qubit case, much higher-order unitary $t$-designs are investigated (See more information in Ref.~\cite{scott2008optimizing}).
In addition, the exact construction of unitary $t$-design for any $t$ is proposed in Ref.~\cite{PRXQuantum.2.030339}.



\subsection{Discretization of the integral}

Next, we approximate $\rho_{n}$ by replacing the integral over $\lambda$ with the sum over $N$ segments in $[0, 1]$. 
Note that in some cases, we need to change the integral variable prior to this discretization, in order to avoid the divergence of the sum.
For example, the Bures measure (\ref{Bures}) diverges at $\lambda=0, 1$, and thereby requires the change of the integral variable for the discretization.

Now suppose that given a one-to-one function $\lambda=g(x)$, the state (\ref{rho_cn}) is rewritten as
\begin{equation}
    \rho_{n}=\frac{1}{|{\cal C}_{n}|}\int_{g^{-1}(0)}^{g^{-1}(1)}dx\, h(x)\sum_{U\in{\cal C}_{n}} (U\Tilde{\Lambda} U^\dagger)^{\otimes n},
\end{equation}
where
\begin{equation}
    h(x)=f(g(x))\frac{dg}{dx}
\end{equation}
and
\begin{equation}
    \Tilde{\Lambda}=g(x)\ket{0}\bra{0}+\left[1-g(x)\right]\ket{1}\bra{1}.
\end{equation}
We then replace the integral with the sum by using the trapezoidal rule
\begin{equation}
    \Delta_i=\frac{|g^{-1}(1)-g^{-1}(0)|}{N}\left(i-\frac{1}{2}\right)
    \label{trapezoidal}
\end{equation}
to find
\begin{equation}
    \Tilde{\rho}_{n}=\frac{1}{N|{\cal C}_{n}|}\sum_{i=1}^Nh(\Delta_i)\sum_{U\in{\cal C}_{n}} (U\Tilde{\Lambda}_i U^\dagger)^{\otimes n},
    \label{rho_approx}
\end{equation}
where we have approximated the density matrix $\Tilde{\Lambda}$ with
\begin{equation}
    \Tilde{\Lambda}_i=w_i(0)|0\rangle\langle 0| + w_i(1)|1\rangle\langle 1|,
\end{equation}
using the probability distribution $w_i(a)$ such that
\begin{equation}
    w_i(0)=g(\Delta_i),
    \qquad
    w_i(1)=1-g(\Delta_i).
\end{equation}
Note that $g(x)=\lambda\in[0,1]$ implies $g(\Delta_i)\in[0,1]$, and we may safely think of $\{w_i(a)\}_{a=0,1}$ as a probability distribution.
Substituting Eq.~(\ref{rho_approx}) to Eq.~(\ref{q_cond}), we then obtain
\begin{equation}
\begin{split}
    &p_{n,k}\approx\\
    &\frac{|{\cal P}_n|\sum_{i=1}^Nh(\Delta_i)\sum_{\Pi\in{\cal P}_{n+1}}\tr(\Tilde{\Lambda}_i\Pi)^{k+1}\tr(\Tilde{\Lambda}_i(\lnot \Pi))^{(n-k)}}{|{\cal P}_{n+1}|\sum_{i=1}^Nh(\Delta_i)\sum_{\Pi\in{\cal P}_n}\tr(\Tilde{\Lambda}_i\Pi)^k\tr(\Tilde{\Lambda}_i(\lnot \Pi))^{(n-k)}}.
    \label{p_approx}
    \end{split}
\end{equation}

For the later convenience, let us show the explicit forms of $g(x)$ employed in the later simulations.
For the flat measure (\ref{flat}), we need no change of the integral variable and thereby set $g(x)=x$, which results in
\begin{equation}
    h(\Delta_i)=w_i(0)=\Delta_i,
\end{equation}
where
\begin{equation}
    \Delta_i=\frac{1}{N}\left(i-\frac{1}{2}\right).
    \label{Delta}
\end{equation}
In contrast, for the Bures measure, we set $g(x)=\sin^2(\pi x/2)$ with $x\in[0,1]$, from which we observe
\begin{equation}
    h(\Delta_i)=\frac{2}{N}\cos^2{(\pi \Delta_i)},
    \quad
    w_i(0)=\sin^2{\left(\frac{\pi }{2}\Delta_i \right)}.
\end{equation}
with Eq.~(\ref{Delta}).

\subsection{Experimental protocols}

The argument presented so far leads to the following experimental protocols of QLLS, which consist of three steps: i) data acquisition, ii) estimation of $p_{n,k}$, and iii) validation of the estimated results.
We hereafter describe the details of each step.


The data acquisition step is composed of the state preparation and measurements.
From the construction of ${\Tilde \rho}_{n+1}$ and Eq.~(\ref{perm}), this step boils down to the pseudo-code in Algorithm \ref{alg}.
Note that this approximation is applicable to other measures $d\mu(\lambda)$ of the mixed state space such as the Hilbert-Schmidt measure \cite{bengtsson_zyczkowski_2006} by appropriately choosing $h(\Delta_i)$ and $w_i(a)$.

\begin{algorithm}[H]
\caption{Subroutine for data acquisition}
\algsetup{indent=2em}
\label{alg}
\begin{algorithmic}
\REQUIRE $M$, $N$, ${\cal C}_{n+1}$, $h(\Delta_i)$, $w_i(a)$, $n+1$ quantum registers.
\FOR{$s=1, 2, \dotsc, M$}
\STATE Draw $\Delta_i$ from the probability distribution $\{h(\Delta_i)\}_{i=1}^N$.
\STATE Calculate $i$ from $\Delta_i$.
\STATE Draw $a$ from the probability distribution $\{w_i(a)\}_{i=0,1}$.
\IF{$a=0$,}
\STATE Prepare $|0\rangle\langle0|$ for all quantum registers.
\ELSIF{$a=1$,}
\STATE Prepare $|1\rangle\langle1|$ for all quantum registers.
\ENDIF
\STATE Draw $U$ from ${\cal C}_{n+1}$ randomly.
\STATE Apply $U$ to all the quantum registers.
\STATE Measure all the quantum registers along the computational basis.
\ENDFOR
\RETURN $M$ quadruplets of $U$, $i$, $a$, and $(n+1)$-bit sequence.
\end{algorithmic}
\end{algorithm}


In the estimation step, we evaluate an estimator $p_{n,k}^{\rm est}$ of the conditional probability $p_{n,k}$ from the classical data obtained in  Algorithm~\ref{alg}.
From the approximation of $p_{n,k}$ developed so far, we may expect two mutually distinct ways of the estimation.

The first way of the estimation (Estimation 1) comes from the fact that $p_{n,k}$ is the conditional probability defined in Eq.~(\ref{q_cond}).
In the procedure, out of the $M$ instances of the $(n+1)$-bit sequences,
we sift those having $k$ heads in the measurements of the $n$ qubits associated with ${\cal H}_1\otimes\dots\otimes{\cal H}_n$.
For these sifted instances, we calculate the relative frequency of the head occurring about the measurement of the $(n+1)$-th qubit as the estimator $p_{n,k}^{\rm est}$.
In contrast, the second way of estimation (Estimation 2) employs Eq.~(\ref{p_approx}) for the calculation of $p_{n,k}$, given $\tr(\Tilde{\Lambda}_i\Pi)$ for all $i$ and $\Pi\in {\cal P}_n$ experimentally.
More precisely, for every $i$, we first estimate $\tr(\Tilde{\Lambda}_i\Pi)$ as the relative frequency of the head occurring in a quantum register associated with ${\cal H}_j$ with $1\le j\le n$, and then calculate $p_{n,k}$ by substituting the estimated value of $\tr(\Tilde{\Lambda}_i\Pi)$ to Eq.~(\ref{p_approx}).



The above estimation procedures can be regarded as the point estimation: Estimation 1 directly evaluates $p_{n,k}$, whereas  Estimation 2 indirectly does so with the use of Eq.~(\ref{p_approx}) as mentioned.
This subtle difference in the estimation procedures could lead us to difficulty in the calculation of the variance or confidence interval of $p_{n,k}^{\rm est}$, since the law of the error propagation we may use in Estimation 2 requires asymptotic normality of the RHS of Eq.~(\ref{p_approx}) as a function of $M$, but this assumption does not necessarily hold in actual experiments \cite{Ku_1966}.

Bearing this feature in mind, we take the following alternative in the validation step.
Suppose that we have repeated the data acquisition step and estimation step $K$ times.
Let $p_{n,k}^{{\rm est}, (l)}$ be the estimator obtained from the $l$-th run of the data acquisition and estimation step.
In the validation step, given $\{p_{n,k}^{{\rm est}, (l)}\}_{l=1}^K$, we evaluate their arithmetic average
\begin{equation}
    \braket{p_{n,k}^{\rm est}}=\frac{1}{K}\sum_{l=1}^{K}p_{n,k}^{{\rm est}, (l)},
\end{equation}
and the variance
\begin{equation}
    \delta p_{n,k}^2=\frac{1}{K}\sum_{l=1}^{K}\left(p_{n,k}^{{\rm est}, (l)}\right)^2-\braket{p_{n,k}^{\rm est}}^2.
\end{equation}
Note that $\braket{p_{n,k}^{\rm est}}$ is also the estimator of $p_{n,k}$.

Although it suffices to evaluate the estimator $\braket{p_{n,k}^{\rm est}}$ and the variance $\delta p_{n,k}^2$ for the validation, it is convenient in practice to employ the mean squared error (MSE)
\begin{equation}
d_{n,k}=\sqrt{(\braket{p_{n,k}^{\rm est}}-p_{n,k})^2 + \delta p_{n,k}^2 },
\end{equation}
since $d_{n,k}$ simultaneously describes the bias $\braket{p_{n,k}^{\rm est}}-p_{n,k}$ and the variance $\delta p_{n,k}^2$ on an equal footing.


The whole procedure of the experimental protocol is given as Algorithm \ref{whole_procedure}.
We will examine its validity in the following section.

\begin{algorithm}[H]
\caption{Implementation of QLLS}
\algsetup{indent=2em}
\label{whole_procedure}
\begin{algorithmic}
\REQUIRE $K$, $M$, $N$, ${\cal C}_{n+1}$, $h(\Delta_i)$, $w_i(a)$, $n+1$ quantum registers.
\STATE Choose Estimation 1 or 2.
\FOR{$l=1, 2, \dotsc, K$}
\STATE Perform Algorithm \ref{alg}.
\STATE Perform the chosen estimation procedure.
\RETURN $p_{n,k}^{{\rm est}, (j)}$.
\ENDFOR
\STATE Evaluate $\braket{p_{n,k}^{\rm est}}$, $\delta p_{n,k}^2$, and $d_{n,k}$.
\RETURN $\braket{p_{n,k}^{\rm est}}$, $\delta p_{n,k}^2$, and $d_{n,k}$.
\end{algorithmic}
\end{algorithm}

\section{Simulation Results and Discussions}
\label{sim}

In Algorithm \ref{whole_procedure}, the estimator $\braket{p_{n,k}^{\rm est}}$ is expected to converge to the corresponding analytical value $p_{n,k}$, as the number of measurements $M$ increases.
On the other hand, how the variance or MSE converges may depend on the estimation procedures. 
In this section, we check whether these expectations hold by using simulation results.

\begin{figure*}
    \centering\includegraphics[width=1.0\textwidth]{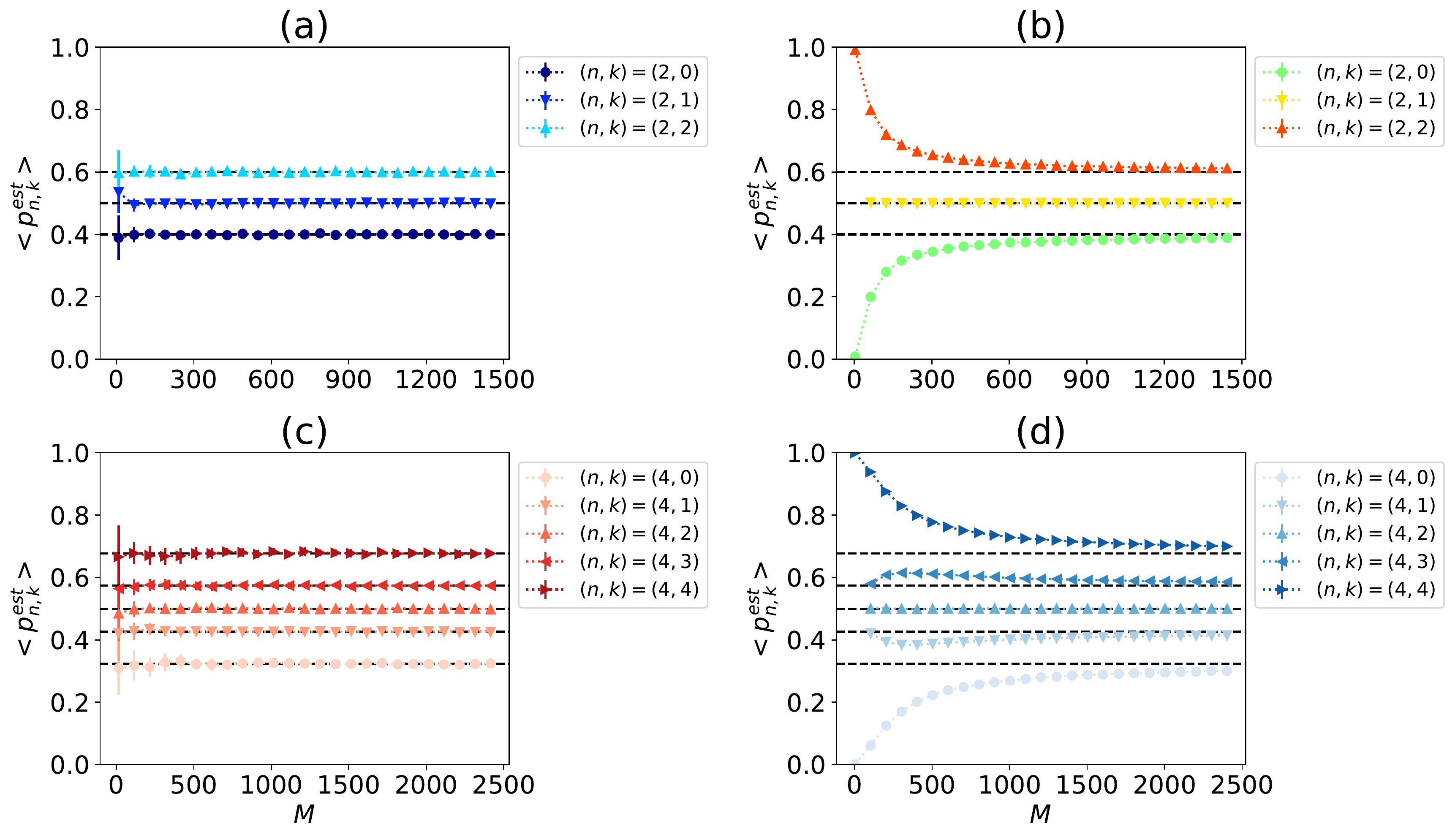}
    \caption{Plots of the estimator $\braket{p_{n,k}^{\rm est}}$ with the flat measure: (a) $n=2$ with Estimation 1, (b) $n=2$ with Estimation 2, (c) $n=4$ with Estimation 1, and (d) $n=4$ with Estimation 2. 
  All the estimators approach the corresponding analytical values $p_{n,k}$ (dotted lines) given in Table \ref{values}, irrespective of the estimation procedure employed.
  The error bars stand for the standard deviation $\delta p_{n,k}^{\rm est}$. This suggests robustness of Algorithm \ref{whole_procedure} with respect to the choice of the post-processing procedures for sufficiently large $M$.}
    \label{flat_convergence}
\end{figure*}
In our simulation of Algorithm~\ref{whole_procedure}, we set $N=50$ to discretize the integral and took the Clifford group and icosahedral group as the unitary $n+1$-design ${\cal C}_{n+1}$ for $n=2,4$, respectively~\cite{PhysRevA.90.030303,PRXQuantum.2.030339} 
(see Ref.~\cite{PhysRevA.90.030303} of the concrete forms of unitary $t$-designs)
.
The corresponding analytical values are summarized in Tab.~\ref{values}.
On the basis of the above setup, we numerically check the convergence of $\braket{p_{n,k}^{\rm est}}$, $\delta p_{n,k}^2$, and $d_{n,k}$ by varying $M$.
We set $K=30$, in order to enumerate the estimator $p_{n,k}^{\rm est}$, its variance $\delta p_{n,k}^2$ and MSE $d_{n,k}$ for every $M$.

\subsection{Flat measure}

Figures~\ref{flat_convergence} shows the behavior of the estimator $\braket{p_{n,k}^{\rm est}}$ against the number of measurements $M$ performed in Algorithm \ref{whole_procedure}. 
From these graphs, we can see that the estimators for $n=2, 4$ converge to the corresponding analytical values as the number of measurements $M$ increases.
Note that the estimator also approximately obeys the relation $\braket{p_{n,k}^{\rm est}}+\braket{p_{n,n-k}^{\rm est}}\approx1$;
the duality relations are robust against the statistical error.
\begin{figure}[htbp]
\centering
\includegraphics[width=0.5\textwidth]{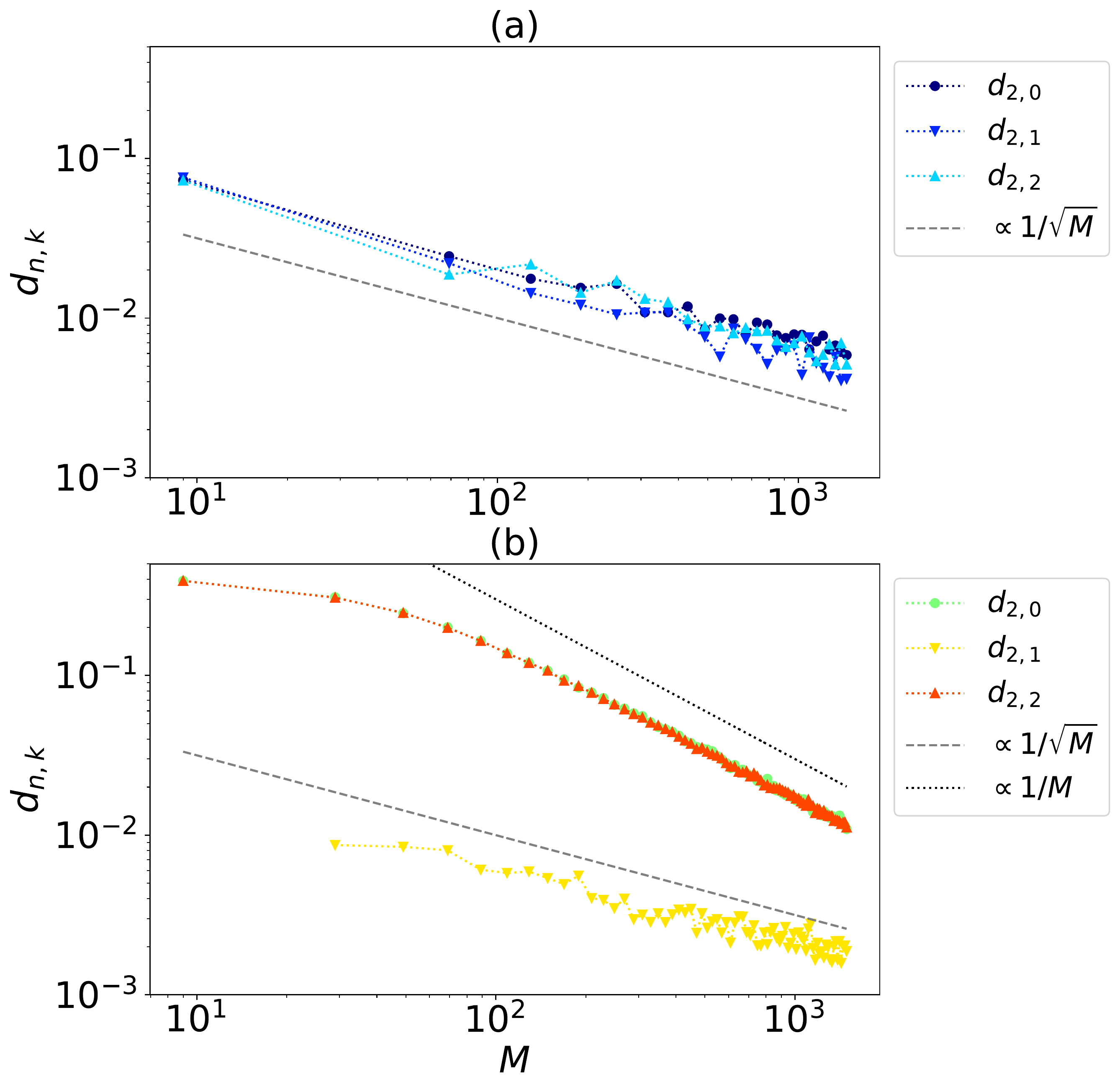}
\caption{Plots of the MSE $d_{n,k}$ for $n=2$ with the flat measure as the functions of $M$: (a) Estimation 1 and (b) Estimation 2. 
  The deviations $d_{2,0}$ and $d_{2,2}$ in Fig.~\ref{Fig:flatn5} (b) behave as ${\cal O}(1/M)$, whereas all the others behave as ${\cal O}(1/\sqrt{M})$, in accordance with the central limit theorem.
  }
  \label{Fig:flatn5}
\end{figure}

On the other hand, how the estimator $\braket{p_{n,k}^{\rm est}}$ converges depends on which estimation procedure we have employed.
For Estimation 1, the estimator quickly approaches the analytical value even at $M={\cal O}(10)$, and the standard deviation $\delta p_{n,k}$ decrease as $M$ increases. 
In contrast, for Estimation 2, the estimator has finite bias $\braket{p_{n,k}^{\rm est}}-p_{n,k}$ up to $M={\cal O}(100)$, which vanishes as $M$ increases.

In order to analyze how MSE depends on $M$, we have plotted MSE $d_{n,k}$ in Figs.~\ref{Fig:flatn5} (a) and (b).
Figure~\ref{Fig:flatn5} (a) shows that the MSE is proportional to $1/\sqrt{M}$ when we employ Estimation 1, implying the consistency with the central limit theorem.
On the other hand, Fig.~\ref{Fig:flatn5} (b) shows that the MSE is proportional to $1/\sqrt{M}$ or $1/M$ in Estimation 2.
Therefore, by putting the exceptional case ($d_{2,0}$ and $d_{2,2}$ for Estimation 2) aside, two estimation procedures are in the same order of the convergence ${\cal O}(1/\sqrt{M})$.

Let us take a closer look at the scaling behavior of MSE $d_{n,m}$ in Estimation 2. 
Figure \ref{flat_convergence} (b) shows that i) there is non-zero bias for $(n,k)=(2,0), (2,2)$ even when $M={\cal O}(10^3)$ whereas no bias for $(n,k)=(2,1)$, and ii) the variance $\delta p_{n,k}^2$ is negligibly small.
Therefore, to the scaling behavior of MSE $d_{n,m}$, the bias is dominant for $(n,k)=(2,0), (2,2)$, whereas the variance is so for $(n,k)=(2,1)$.
%
From the above observations, we find that the standard error (the square root of the variance $\delta p_{n,k}^2$) behaves as ${\cal O}(1/\sqrt{M})$, showing its consistency with the central limit theorem.
On the other hand, the scaling with ${\cal O}(1/M)$ for $(n,k)=(2,0), (2,2)$ comes from the squared bias term $(\braket{p_{n,k}^{\rm est}}-p_{n,k})^2$.

\subsection{Bures measure}
To examine the robustness of our experimental protocols with respect to the choice of the probability distribution $h(x)$, we have considered the conditional probability $p_{n,k}$ for the Bures measure.
Similarly to the case of the flat measure, the conditional probability converges to its corresponding analytical value as the number of measurements $M$ increases.
This shows the robustness of our protocols.
See Appendix \ref{Bures_figures} for the details of the simulation results.

\section{Conclusion}
\label{conc}
In this work, we have proposed QLLS and its implementation methods.
The degrees of freedom with respect to unitary transformations are responsible for the non-trivial overall factor.
On the basis of this fact, we may say that QLLS is a quantum effect for the few-body systems manifesting itself through the conditioned post-processing procedures.

In machine learning community, LLS has been employed in naive Bayes classifier in order to improve the accuracy of the estimation of the parameters characterizing the classes \cite{10.5555/3041838.3041916}.
The QLLS will be useful for quantum naive Bayes classifier \cite{Shao_2020} to improve its accuracy.

In our simulations, we have considered two estimation procedures. 
Both the procedures attained $d_{n,k}={\cal O}(10^{-3})$ when $M\approx10^3$. 
By taking into account that we have repeated the procedures $K=30$ times in our simulations, this suggests that we can perform actual experiments with good precision for $n+1=3$ qubits with $KM\approx3\times10^4$ measurements along the computational basis.
Note that this number of measurements could be reduced for Estimation 1 if we use $p_{n,k}^{\rm est}$ as its estimator and calculate the confidence interval by using the formula $\sqrt{p_{n,k}^{\rm est}(1-p_{n,k}^{\rm est})/M}$. 
Then we may put $K=1$.
In contrast, we cannot expect such a reduction for Estimation 2, since the calculation of the confidence interval requires the use of the error propagation formula, whose promise does not necessarily hold.

Our simulations can be easily extended for the cases of the general $n$ by choosing appropriate unitary $t$-designs: 
We have employed two unitary groups (Clifford group and icosahedral group), which are already realized in some experiments using superconducting qubits~\cite{PhysRevA.90.030303,PRXQuantum.2.030339}.
Moreover, as announced before, many other unitary  $t$-design are known in previous studies.
Particularly in the $SU(2)$ case, much higher-order unitary $t$-designs are investigated (See more information in Ref.~\cite{scott2008optimizing}).
In addition, the exact construction of unitary $t$-design for any $t$ is proposed in Ref.~\cite{PRXQuantum.2.030339}.

Furthermore, we briefly explicate the connection between our study and classical shadow~\cite{Huang_2020,PRXQuantum.2.030348,Elben_2022,nakaji2022measurement}. Classical shadow is an efficient measurement protocol that estimates multiple observables simultaneously by utilizing random Clifford circuits (unitary 3-design) measurements and classical post-processing techniques instead of Haar random unitaries.The random Clifford circuits are comprised of two schemes: $n$-qubit random Clifford circuits or single-qubit random Clifford ones. The latter scheme is tantamount to our estimation methodologies, because our methodologies perform random measurements based on unitary $t$-designs in lieu of Haar random unitaries. Despite this similarity, our protocol simultaneously estimates the probabilities for all the possible occurring outcomes with fixed measurement basis, whereas the classical shadow algorithm can estimate the expectation values of the simultaneously measurable observables with possible inefficiencies due to varying the measurement basis.
For this difference, our algorithm is useful not for the expectation values but for the estimation of the probabilities;
our algorithm is complementary to the classical shadow and expands the range of the applications of random measure measurements with post-processing.

Note that the estimation procedures are not necessarily limited to the two algorithms examined in this work.
For example, in order to reduce the number of measurements for the calculation of Eq.~(\ref{p_approx}), we may reuse measurement outcomes to estimate the probabilities $|\bra{0}U\ket{0}|^2$ and $|\bra{1}U\ket{1}|^2$ for each $U\in{\cal C}_{n+1}$, which appear in $\tr[\tilde{\Lambda}_i\Pi]$. 
We could perform this alternative procedure at the cost of the simplicity of the total variance formula, since the covariance between different segments $\Lambda_i$ may arise in this case.

The remaining issue is the actual experimental demonstration of QLLS using existing quantum computers or other physical systems.
On the ground of the preliminary simulation results in this work, it may require experimental techniques to suppress inevitable noise in the quantum computers such as quantum error mitigation~\cite{PhysRevLett.119.180509,PhysRevX.8.031027,endo2021hybrid}.
Such detailed analyses are out of the scope of this work, but we are planning to perform such experiments, whose outcomes will be reported elsewhere.


\begin{acknowledgments}
This work was supported by MEXT Quantum Leap Flagship Program (MEXT Q-LEAP) Grant Number JPMXS0120319794.
\end{acknowledgments}

\appendix
\section{Non-zero global quantum discord of $\rho_2$ for the flat measure}
\label{discord_cal}
It is of interest to ask why the QLLS has the non-trivial overall factor.
In Sec.~\ref{QLLS}, we have attributed this overall factor to the unitary degrees of freedom inherent in QM.
Indeed, for the state $\rho_2$ with the flat measure, the integration of the unitary degrees of freedom leads to a quantum correlation called global quantum discord \cite{PhysRevA.84.042109}, whose measure is given by
\begin{align}
    \mathcal{D}(\rho_n)=\min_{\{\Pi_j\}}\left[S(\rho_n||\Phi(\rho_n))-\sum_{j=1}^nS(\rho_n^{(j)} ||\Phi_j(\rho_n^{(j)})) \right].
\end{align}
Here, $S(\rho||\sigma)=\mathrm{tr}[\rho\log{\rho}-\rho\log{\sigma}]$ is the quantum relative entropy, $\Pi_j$ is a projection operator acting on ${\cal H}_j$, and $\rho_n^{(j)}$ is the reduced density operator on the $j$-th qubit. $\Phi(\rho_n)$ and $\Phi_j(\rho_n^{(j)})$ are the post-measurement states for the total state $\rho_n$ and local state $\rho_n^{(j)}$, respectively.
Note that the $\mathcal{D}(\rho_n)$ is zero for the classical state.

For the calculation of the global quantum discord of the $\rho_2$ with the flat measure, we use the spectral decomposition (\ref{dec}) and the Euler angle parametrization of $SU(2)$ group.
Performing the integral, we obtain
\begin{align}
    \rho_2=\frac{1}{4}\left(\openone^{\otimes2}+\frac{1}{3}\sigma_x^{\otimes2}+\frac{1}{3}\sigma_y^{\otimes2}+\frac{1}{9}\sigma_z^{\otimes2} \right),
\end{align}
where $\sigma_x$, $\sigma_y$, and $\sigma_z$ are the Pauli matrices.
For this case, an analytical formula of the global quantum discord has been known \cite{xu2013analytical}.
By using the formula, we find
\begin{align}
    D(\rho_2)=1-\frac{5}{3}\ln2-\ln3 +\frac{5}{9}\ln5+\frac{7}{18}\ln7 \simeq 0.397.
\end{align}
Note that $\rho_2$ is not entangled from Eq.~(\ref{def_rho}).
Therefore, $\rho_2$ for the flat measure is a separable state with non-zero global quantum discord, from which the non-trivial overall factor of QLLS arises.

\section{Proof of Eq.~(\ref{reciprocal})}
\label{proof_reciprocal}

Suppose that our measure obeys the symmetry (\ref{f_sym}). 
Let us transform the integration variable $\lambda$ in Eq.~(\ref{def_rho}) with 
\begin{equation}
    \lambda^\prime=1-\lambda.
\end{equation}
Then, it follows from Eq.~(\ref{f_sym}) that
\begin{equation}
    f(\lambda)=f(\lambda^\prime).
\end{equation}
To make it explicit that $\Lambda$ depends on $\lambda$, hereafter we write $\Lambda$ as $\Lambda(\lambda)$.
We then immediately observe
\begin{equation}
    \Lambda(\lambda)=\openone-\Lambda(\lambda^\prime),
\end{equation}
which implies
\begin{equation}
    \tr(\rho P)=1-\tr(U\Lambda(\lambda^\prime)U^\dagger P).
\end{equation}
Therefore, we find
\begin{align}
    \tr(\rho_n P_{n,k})
    &=\int d\nu(U)\int_0^1df(\lambda)\tr(\rho P)^k[1-\tr(\rho P)]^{n-k}
    \nonumber\\
    &=\int d\nu(U)\int_0^1df(\lambda^\prime)
    \Big\{[1-\tr(U\Lambda(\lambda^\prime)U^\dagger P)]^k\nonumber\\
    &\qquad\qquad\times\tr(U\Lambda(\lambda^\prime)U^\dagger P)^{n-k}\Big\}
    \nonumber\\
    &=\int d\nu(U)\int_0^1df(\lambda)[1-\tr(\rho P)]^k\tr(\rho P)^{n-k}
    \nonumber\\
    &=\tr(\rho_n P_{n,n-k}),
\end{align}
which leads to
\begin{equation}
    \tr(\rho_{n+1}P_{n+1,n-k+1})=\tr(\rho_{n+1}P_{n+1,k}).
\end{equation}
On the other hand, we have
\begin{equation}
    P_{n+1,k+1}+P_{n+1,k}=P^{\otimes k}\otimes\openone\otimes(\lnot P)^{\otimes (n-k)},
\end{equation}
from which we obtain
\begin{align}
    &\tr(\rho_{n+1} P_{n+1,k+1})+\tr(\rho_{n+1} P_{n+1,n-k+1})
    \nonumber\\
    &=\tr(\rho_{n+1} P^{\otimes k}\otimes\openone\otimes(\lnot P)^{\otimes (n-k)})
    \nonumber\\
    &=\int d\mu(\rho)\tr(\rho P)^k[1-\tr(\rho P)]^{n-k}
    \nonumber\\
    &=\tr(\rho_n P_{n,k}).
\end{align}
We then arrive at
\begin{align}
    &p_{n,k}+p_{n,n-k}
    \nonumber\\
    &=\frac{\tr(\rho_{n+1} P_{n+1,k+1})}{\tr(\rho_n P_{n,k})}
    +\frac{\tr(\rho_{n+1} P_{n+1,n-k+1})}{\tr(\rho_n P_{n,n-k})}
    \nonumber\\
    &=\frac{\tr(\rho_{n+1} P_{n+1,k+1})+\tr(\rho_{n+1} P_{n+1,n-k+1})}{\tr(\rho_n P_{n,k})}
    \nonumber\\
    &=\frac{\tr(\rho_n P_{n,k})}{\tr(\rho_n P_{n,k})}
    \nonumber\\
    &=1,
\end{align}
which completes the proof.

\section{QLLS for the measure (\ref{classical})}
\label{derivation_cl}

To derive QLLS, let us note that the concrete expression of $\rho_n$ for the measure (\ref{classical}) is given by
\begin{equation}
    \rho_n=\int_0^1 d\lambda\Lambda^{\otimes n}
\end{equation}
from the spectral decomposition (\ref{dec}).
We then find
\begin{equation}
    \tr(\rho_nP_{n,k})=\int_0^1 d\lambda \lambda^k(1-\lambda)^{n-k}=B(n-k+1,k+1), 
\end{equation}
where $B(n,k)$ is the Beta function.
Substituting this to Eq.~(\ref{q_cond}), we find
\begin{equation}
    p_{n,k}=\frac{B(n-k+1,k+2)}{B(n-k+1,k+1)}=\frac{k+1}{n+2}=p_{n,k}^{\rm cl},
\end{equation}
since $B(n, k) = (n-1)!(k-1)!/(n+k-1)!$.

\begin{figure*}
    \centering
    \includegraphics[width=1.0\textwidth]{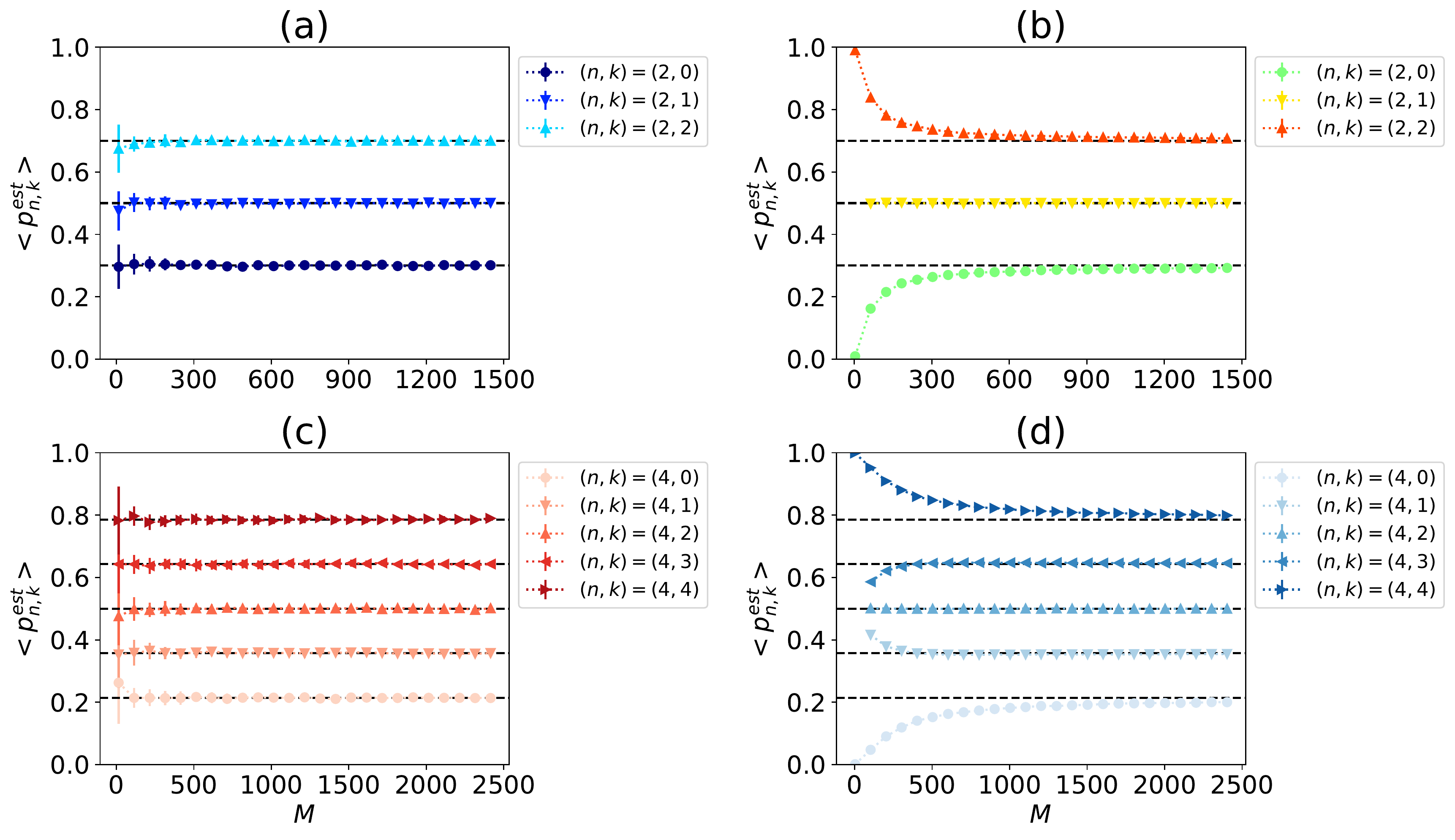}
    \caption{Plots of the estimator $\braket{p_{n,k}^{\rm est}}$ with the Bures measure: (a) $n=2$ with Estimation 1, (b) $n=2$ with Estimation 2, (c) $n=4$ with Estimation 1, and (d) $n=4$ with Estimation 2. 
  All the estimators approach the corresponding analytical values $p_{n,k}$ (dotted lines) given in Table \ref{values}, irrespective of the estimation procedure employed.
  The error bars stand for the standard deviation $\delta p_{n,k}^{\rm est}$. This suggests robustness of Algorithm \ref{whole_procedure} with respect to the choice of the measures to be implemented. 
  }
\label{Fig:Bures}
\end{figure*}
\section{Simulation results on the Bures measure}
\label{Bures_figures}

Figures \ref{Fig:Bures} show the simulation results of the QLLS for the Bures measure.
Similarly to the flat measure, all the experimental values approach the corresponding analytical values as the number of measurements $M$ increases.
The duality relation $\braket{p_{n,k}^{\rm est}}+\braket{p_{n,n-k}^{\rm est}}\approx1$ also approximately holds.
%


\bibliography{apssamp}

\end{document}